\documentclass[dvips]{article}
\usepackage{icrctc07}

\title{A kinetic approach to non resonant modes and growth rates of streaming
instability: consequences for shock acceleration}
\shorttitle{Growth rates for streaming instability}

\authors{Pasquale Blasi$^{1}$, Elena Amato$^{1}$}
\shortauthors{Blasi and Amato}
\afiliations{$^1$INAF/Osservatorio Astrofisico di Arcetri\\
Largo E. Fermi 5, 50126 Firenze (Italy)}
\email{blasi@arcetri.astro.it}

\abstract{We show here that a purely kinetic approach to the
  excitation of waves by cosmic rays in the vicinity of a shock front
  leads to predict the appearance of a non-alfvenic fastly growing
  mode which is the same that was found by Bell (2004) by treating the
  plasma in the MHD approximation. The kinetic approach we present is
  more powerful in that it allows us to investigate different models
  for the compensation of the cosmic ray current in the background
  upstream plasma.}

\begin{document}
\maketitle

\section{Introduction}

There are different perspectives from which we can look at the role of
supernova remnants (SNRs) as sources of the bulk of cosmic rays: from
one, they represent the only class of sources which are energetically
viable in terms of hosting the necessary energy, provided about
$10-20\%$ of the kinetic energy of the expanding shell is converted
into accelerated articles. On the other hand it has been known for a
long time \cite{lagage83a,lagage83b} that the standard model needs to
be pushed to its limits and beyond in order to explain the maximum
energies which are observed in cosmic rays (about $10^6$ GeV for
protons). Observationally there have been recent claims of detection
of strong magnetic fields, as measured from the thickness of the X-ray
brightness profiles of several SNRs. These fields would make the
acceleration process certainly easier if they rearrange topologically
in a way to scatter the particles approximately at the Bohm rate. Both
the issues of efficient acceleration and nonlinear amplification of
the magnetic field lead to the need to develop a nonlinear theory of
particle acceleration at SNR shocks (see \cite{blasiicrc}, these
proceedings). 

From the theoretical point of view, the most important aspect consists
of providing a fundamental explanation of the mechanisms that relate
cosmic rays as accelerated particles to the magnetic field
amplification which has been observed. The natural channel seems to be
streaming instability, but other mechanisms are also possible. 

Here we discuss a kinetic approach to waves excited by the streaming
of cosmic rays upstream of the shock and show that there are modes
which may be responsible for magnetic field amplification by a factor
of order $\sim 100-1000$ for typical parameters of a SNR. A similar
conclusion was also obtained by using a different approach in
\cite{bell04}.

\section{Streaming instability and compensating currents}

In the reference frame of the shock, cosmic rays are approximatively
stationary and roughly isotropic. The upstream background plasma moves
with a velocity $v_s$ towards the shock and is made of protons and
electrons. The charge of cosmic cays, assumed to be all protons
(positive charges) is compensated by processes which depend on the
microphysics and need to be investigated accurately. For instance, if
$n_i$ and $n_e$ are the proton and electron density in the background
plasma respectively, then one can compensate the cosmic ray charge
$N_{CR}e$ by either requiring that $n_i+N_{CR}=n_e$, thereby assuming
that the electrons and protons have different densities (and different
velocity with respect to the shock), or we can require that protons and
electrons in the background gas have the same density and velocity and
that there is a fourth population of particles, which are cold
electrons, at rest in the shock frame which drift together with cosmic
rays and cancel their positive charge. The first avenue was adopted in
\cite{achter83}, while the second was used in \cite{zweib78,zweib03}.  
The latter option resembles more closely the underlying physical
approach of \cite{bell04} where the background gas was however treated
in the MHD approximation, and as such described in terms of a single
density ($n_i=n_e$). In the reference frame in which cosmic rays are
isotropic (namely $\partial f_{CR}/\partial \mu=0$, where $f_{CR}$ is
the distribution function of cosmic rays) the distribution functions
of the background ions, electrons and cold electrons are respectively
\begin{eqnarray*}
f_{i}(p)&=&\frac{n_i}{p^2}\delta (p-m_i v_s) \delta (\mu+1) \\
f_{e}(p)&=&\frac{n_e}{p^2}\delta (p-m_e v_s) \delta (\mu+1) \\
f_{e}^{cold}(p)&=&\frac{N_{CR}}{2p^2} \delta (p). 
\end{eqnarray*}
The cosmic ray distribution can also be written as
\begin{equation}
f_{CR}(p)=\frac{N_{CR}}{2} g(p),
\end{equation}
where $g(p)$ is a function normalized so that $\int_0^{\infty} dp~p^2
g(p)=1$. The dispersion relation of waves in this composite plasma can
be written as \cite{krall}:
\begin{eqnarray*}
\frac{c^2 k^2}{\omega^2}=1+\sum_\alpha \frac{2\pi q_\alpha^2}{\omega}
\int_0^{\infty} dp \int_{-1}^{+1} d\mu \\
\frac{p^2 v(p) (1-\mu^2)}{\omega-k v(p) \mu \pm \Omega_\alpha}\left[
\frac{\partial f_\alpha}{\partial p}+\left( \frac{k
  v}{\omega}-\omega\right)\frac{1}{p} \frac{\partial f_\alpha}{\partial \mu}
\right],
\label{eq:disp}
\end{eqnarray*} 
where $\alpha$ runs over the particle species in the plasma, $\omega$
is the wave frequency corresponding to the wavenumber $k$ and
$\Omega_\alpha$ is the relativistic gyrofrequency of the particles of
type $\alpha$, which in terms of the cyclotron frequency and the
Lorentz factor is $\Omega_\alpha=\Omega_\alpha^*/\gamma$. For the
background plasma and for the cold electrons compensating the cosmic
ray charge one has $\Omega_\alpha\approx\Omega_\alpha^*$. 

Carrying out the usual algebraic handling of the different terms
appearing in Eq. \ref{eq:disp} for the four components of the plasma,
one obtains the following dispersion relation, as written in the shock
frame: 

\begin{eqnarray*}
\frac{c^2 k^2}{\omega^2}=1-\frac{4\pi e^2 n_i}{\omega^2}(\omega+k
v_s)\times\\
\left[ \frac{1}{m_p}\frac{1}{\omega+k v_s \pm\Omega_i^*} + 
\frac{1}{m_e}\frac{1}{\omega+k v_s \pm\Omega_e^*}\right]\\
-\frac{4\pi e^2 N_{CR}}{\omega m_e}\frac{1}{\omega\pm \Omega_e^*}+
\frac{4\pi e^2 N_{CR}}{\omega} \times\\
\left\{
\frac{1}{4k}\int_0^\infty \frac{dg}{dp}\left[\pm 2 p_k p +
  (p_k^2-p^2)\ln |\frac{1\mp p/p_k}{1\pm p/p_k}|\right]-\right.\\
\left.-\frac{i\pi}{4k}\int_{Max(p_k,p_0)}^\infty dp \frac{dg}{dp}(p^2-p_k^2)
\right\}.
\end{eqnarray*}

Here we have introduced the momentum $p_k=m_p \Omega_i^*/k$ which is
the minimum momentum of the particles which can resonate with a wave
with wavenumber $k$. The momentum $p_0$ is the minimum momentum of the
accelerated particles. The above dispersion relation is considerably
simplified in the limit of frequency $\omega$ small compared with
$\Omega_i^*$. In particular we shall assume that $\omega+k v_s \ll
\Omega_i^* \ll |\Omega_e^*|$. Moreover we will neglect the
displacement current (the unity), so that the dispersion equation
reads
\begin{eqnarray*}
v_A^2 k^2 = (\omega+k v_s)^2 \pm \omega \Omega_i^* \frac{N_{CR}}{n_i}
\left\{ 1 \pm \right.\\
\left.\frac{p_k}{4} \int_{p_0}^\infty dp \frac{dg}{dp} 
\left[\pm 2 p_k p + (p_k^2-p^2)\ln |\frac{1\mp p/p_k}{1\pm
    p/p_k}|\right]\right.\\
\left.\mp \frac{i\pi}{4}p_k\int_{Max(p_k,p_0)}^\infty dp
\frac{dg}{dp}(p^2-p_k^2)\right\}. 
\end{eqnarray*}

The dispersion relation in the frame of the upstream plasma is
obtained by the change $\omega\to \omega-k v_s$. Moreover, neglecting
again $\omega$ with respect to $k v_s$ we get 

\begin{equation}
v_A^2 k^2 = \omega^2 \mp \frac{k v_s \Omega_i^* N_{CR}}{n_i} 
\left[ 1\pm I_1(k) \pm i I_2(k) \right].  
\label{eq:disp1}
\end{equation}

This dispersion relation is basically the same as that obtained in
\cite{bell04} and has therefore the same implications. The terms
proportional to $I_2$ are those describing the resonant interaction of
the waves and particles. The non resonant part is proportional to
$1+I_1$. 

In the following we investigate these implications in the case of a
power law spectrum of cosmic rays $g(p)\propto (p/p_0)^{-4}$. In order
to have the correct normalization of $g(p)$ we must require
$g(p)=(1/p_0^3)(p/p_0)^{-4}$. It follows that the energy density in
the form of accelerated particles is 
\begin{equation}
U_{CR} = \frac{N_{CR}}{2} \int_{p_0}^{p_{max}} dp p^3 c g(p) =
\frac{N_{CR}}{2}p_0 c \ln R
\end{equation}
where $R=p_{max}/p_0$, and $p_{max}$ is the maximum momentum of the
accelerated particles. The cosmic ray number density can then written
in terms of energy density as
\begin{equation}
N_{CR} = \frac{2U_{CR}}{p_0 c \ln R}.
\end{equation}

The constant in front of the square brackets in Eq. \ref{eq:disp1} can
therefore be written as
\begin{equation}
 \frac{k v_s \Omega_i^* N_{CR}}{n_i} = \zeta \frac{k}{r_{L,0}} v_s^2
 \equiv \sigma (k),
\label{eq:def}
\end{equation}
where $\zeta=2\eta (v_s/c)(1/\ln R)$ and $\eta=U_{CR}/(\rho u^2)$ is
the fraction of the kinetic energy that is transformed into cosmic
rays at the shock. $r_{L,0}$ is the gyration radius of protons with
momentum $p_0$. 

The solution for the imaginary part of the frequency from the
dispersion relation Eq. \ref{eq:disp1} can be easily found to be: 
\begin{eqnarray*}
\omega_I\equiv Im[\omega] = -\frac{1}{2}\left( v_A^2 k^2 \pm \sigma (1\pm I_1)
\right) \pm \\
\frac{1}{2} \sqrt{ \left( v_A^2 k^2 \pm \sigma (1\pm I_1)
\right)^2 + \sigma^2 I_2^2}
\label{eq:sol}
\end{eqnarray*}

In Eq. \ref{eq:disp1}, for $k\sim 1/r_{L,0}$, the two functions
$I_1(k)$ and $I_2(k)$ are of order unity. Therefore it
is interesting to investigate what happens in this situation when
$\sigma\ll k^2 v_A^2$ and $\sigma\gg k^2 v_A^2$. 

When $\sigma\ll k^2 v_A^2$ we can expand Eq. \ref{eq:sol} in the ratio
$\sigma/(v_A^2 k^2)\ll 1$ and we get:
\begin{equation}
(\omega_I)_{res} = \frac{\sigma I_2}{2 v_A k}.
\end{equation}

This is the standard result already obtained in previous literature,
which implies the (resonant) growth of Alfven waves weakly driven by
the streaming of cosmic rays. It is however instructive to see what
the condition $\sigma\ll k^2 v_A^2$ corresponds to for typical
parameters of a supernova. Let us adopt $B_0=1\mu G$, $v_s=5000$ km/s,
$n=1~\rm cm^{-3}$, $p_{max}=10^6$ GeV/c, $p_0=1$ GeV/c. The condition
therefore leads to $\eta \ll 10^{-6}$. This is a very stringent
condition, requiring that only a millionth of the SNR energy is
converted into cosmic rays. If this were the case, SNRs would have a
negligible role as sources of galactic cosmic rays. It is possible to
change the values of the parameters and bring the limit on $\eta$ up
to $\sim 10^{-3}-10^{-4}$, but still it appears that the condition
necessary for the growth to be considered as weakly driven by the
cosmic ray streaming is too stringent. In other words, the request
that $\eta\sim 0.1$ typically formulated, naturally leads to the
conclusion that the wave growth takes place in the opposite regime, 
$\sigma\gg k^2 v_A^2$. 
\begin{figure}
\begin{center}
\noindent
\includegraphics [width=0.45\textwidth]{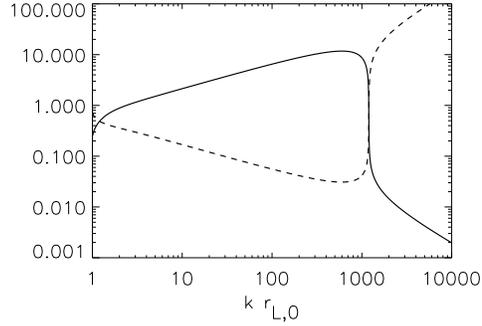}
\end{center}
\caption{Imaginary part (solid curve) and real part (dashed curve) of
  the frequency.}
\label{fig:disp}
\end{figure}

\begin{figure}
\begin{center}
\noindent
\includegraphics [width=0.45\textwidth]{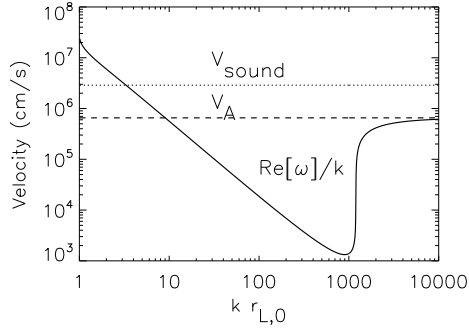}
\end{center}
\caption{Phase speed of the waves compared with the Alfven speed and
  the sound speed.}
\label{fig:speed}
\end{figure}

In this case it is straightforward to show that $\omega_I\approx
\sqrt{\sigma}$.

There are two interesting aspects of this growing mode that deserve
some more attention. As already pointed out in \cite{bell04}, the fact
that $\sigma \gg k^2 v_A^2$ implies that the mode does not resemble an
Alfven wave growing under the effect of cosmic rays. Another point is
that the growth takes place also due to the non-resonant interaction of
the wave with the streaming cosmic rays. 

A more detailed analysis of the modes requires to keep track of the
wavenumber $k$. The imaginary and real parts of the frequency are
shown in Fig. \ref{fig:disp} for $B_0=3\mu G$, $V_s=10^9\rm
cm~s^{-1}$, $\eta=0.1$, $n_i=1 \rm cm^{-3}$. On the x-axis we plotted
$k r_{L,0}$, while the y-axis is in units of $v_s^2/(c r_{L,0})$. 
The plot is limited to the most interesting region $k r_{L,0}>1$. 
The peak in $Im[\omega]$ corresponds to the maximum
growth rate. Note that the highest k part of the plot corresponds to
the case of resonant scattering, and in fact for lower values of the
shock speed (already for $v_s=10^8\rm cm~s^{-1}$) this is the only
mode left. This suggests that at different stages in the evolution of
the remnant the non-resonant or the resonant growth will be relevant. 
Note also that the maximum growth occurs at $k\gg 1/r_{L,0}$, namely
on scales much smaller than the Larmor radius of the particles. The
effect of these magnetic oscillations on the motion of the particles
need to be carefully investigated. 

One final note is on the phase velocity of the waves: in
Fig. \ref{fig:speed} we show the phase velocity $Re[\omega]/k$ as
compared with the Alfven speed and the sound speed (for a temperature
of the upstream gas $T=10^5 K$). We can see that there is a range of
values of $k$ (not the ones where the maximum growth occurs) for which 
the phase speed is larger than the Alfven speed and for small $k$'s
even larger than the sound speed. These supersonic modes are likely to
develop shocks in the precursor that might inhibit  the acceleration
process, also due to the fact that the effective speed of the
scattering centers, $v_s-Re[\omega]/k$ becomes small enough that the
acceleration may be substantially suppressed. 

\section{Conclusions}

We showed that the growing mode found in \cite{bell04} also results
from a kinetic approach. The dispersion relation leading to the
identification of the resonant and non-resonant modes is basically the
same in the two approaches. It is also shown that for typical
parameters of a supernova the magnetic field is likely to be
substantially amplified in the shock vicinity as due to non-resonant
growth of a mode that does not resemble an Alfven wave weakly modified
by cosmic ray streaming. The growth of this mode is faster than the
standard resonant growth obtained in the limit of small efficiency of
acceleration and may make particle acceleration to high energies easier.

\end{document}